# The Heats of Reactions. Calorimetry and Van't-Hoff. 2

I. A. Stepanov

University of Latvia, Rainis bulv. 19, Riga, LV-1586, Latvia

## Abstract

Earlier it has been supposed that the law of conservation of energy for chemical reactions has the following form:

$$dU=dQ-PdV+\sum_i \mu_i dN_i$$

In [1-6] it has been shown that for the biggest part of reactions it must have the following form:

$$dU=dQ+PdV+\sum_i \mu_i dN_i$$

In the present paper this result is confirmed by other experiments.

## 1. Introduction

For chemical processes the law of conservation of energy is written in the following form:

$$dU=dQ-PdV+\sum_i \mu_i dN_i \qquad (1)$$

where dQ is the heat of reaction, dU is the change in the internal energy, $\mu_i$ are chemical potentials and $dN_i$ are the changes in the number of moles.

In [1-6] it has been shown that the energy balance in the form of (1) for the biggest part of the chemical reactions is not correct. In the biggest part of the chemical reactions the law of conservation of energy must have the following form:

$$dU=dQ+PdV+\sum_i \mu_i dN_i \qquad (2)$$

In [1-6] the following method of combining differential equations was used. If there is an equality $df=\pm\alpha dx\pm\beta dy\pm\gamma dz$ ($\alpha$, $\beta$ and $\gamma>0$) and one does not know what signs must there be in front of $\alpha$, $\beta$ and $\gamma$ one proceeds as following. Let $df<0$, if $dx<0$ then $+\alpha$. The signs before $\beta$ and $\gamma$ are found analogically. Here $dQ=dQ_1+dQ_2$, $dQ_1=dU$, $dQ_2=PdV$. Of course, if $df=\pm\alpha dx\pm\beta dy\pm\gamma dz$ is a pure mathematical object, this method is wrong. But for physical processes it is always correct. The sense is the following one: one introduces the quantity of heat dQ in the system and uses for that 100 kg fuel. 80 kg fuel is used for dU and 20 kg is used for PdV. Another solution is: 120 kg for dU and -20 kg for PdV. It has no physical sense.

If there is no chemical reaction then in (2) $\sum_i \mu_i dN_i=0$ and $dQ=dU=PdV=0$. It is impossible to connect (2) and 5) in the limiting case. Eq. (2) describes a chemical phenomenon but (5) describes a physical phenomenon. Chemical phenomena can not be reduced to physical ones. Some authors think that $\sum_i \mu_i dN_i=0$ for any chemical process in equilibrium but it is not true. Discussion of that will be below.

The Van't-Hoff equation is the following one:

$$\partial \ln K/\partial T=\Delta H^0/RT^2 \qquad (3)$$

where K is the reaction equilibrium constant and $\Delta H^0$ is the enthalpy. According to thermodynamics, the Van't-Hoff equation must give the same results as calorimetry because it is derived from the 1st and the 2nd law of thermodynamics without simplifications. However, there is a paradox: the heat of chemical reactions, that of dilution of liquids and that of other chemical processes measured by calorimetry and by the Van't-Hoff equation differ significantly [1-6]. The difference is far beyond the error limits. The reason is that in the derivation of the Van't-Hoff equation it is necessary to take into account the law of conservation in the form of (2), not of (1) [1-6].

If to derive the Van't-Hoff equation using (2), the result will be the following one:

$$\partial \ln K/\partial T=\Delta H^{0*}/RT^2 \qquad (4)$$

where $\Delta H^{0*}=\Delta Q^0+P\Delta V^0$.

The Van't-Hoff equation (4) was often used for determination of the heat of reaction. It gives correct results for reactions with $\Delta V\to 0$. Therefore, it was assumed that for chemical reactions

$$\Delta Q=\Delta H=\Delta U+P\Delta V \qquad (5)$$

However from (1) and (5) it follows

$$\Delta Q=\Delta Q+\sum \mu_i\Delta N_i \qquad (6)$$

This conclusion is absurd: one can not neglect the last term in (6). Some authors [7] suppose that $\sum_i \mu_i dN_i=0$ for any chemical process in equilibrium. It is not true because the change in the Gibbs energy $\Delta G^0$ when reactants turn to the products completely is $\sum_i \mu_i dN_i\neq 0$.

Therefore, (3) must not give correct results neither for $\Delta V\to 0$ nor for $\Delta V\neq 0$. For reactions with $\Delta V\neq 0$, the Van't-Hoff equation gives wrong results. The present theory explains this paradox: for chemical processes (4) must be used.

In the present paper the following processes were investigated: $Al_2Br_6(g)=2AlBr_3(g)$, $AlBr_3(l)=AlBr_3(g)$, $2AlBr_3(l)= Al_2Br_6(g)$, $2V_2O_5(s)=V_4O_{10}(g)$, $2SnO_2(s)=2SnO(g)+O_2$, $2HgO(g)=2Hg(g)+O_2$, $SiO_2(s)=SiO(g)+O$, $Hg(g)+Te(l)=HgTe(s)$, $2Hg(g)+Te_2(g)=2HgTe(s)$.

## 2. Experiments

In [8] the following reaction was considered:

$$Al_2Br_6(g)=2AlBr_3(g) \quad (7)$$

$$\ln K=-14077{,}0/T+147{,}5 \qquad 500\le T\le 900K \quad (8)$$

In [10] the following processes were considered|:

$$AlBr_3(l)= AlBr_3(g) \quad (9)$$

$$2AlBr_3(l)= Al_2Br_6(g) \quad (10)$$

The 1st process is a phase transition, the 2nd one is a chemical reaction. Their heats are given in Table 1. One sees that the heat of evaporation measured by the Van't-Hoff equation is much closer to the experiment than the heat of the reaction. Evaporation is a physical process, it obeys the traditional thermodynamics. The second process obeys the present theory.

In [11, 13-15, 17, 18] the following reactions have been studied:

$2V_2O_5(s)=V_4O_{10}(g)$ [11] (11)

$2SnO_2(s)=2SnO(g)+O_2$ [13] (12)

$2HgO(g)=2Hg(g)+O_2$ [14] (13)

$SiO_2(s)=SiO(g)+O$ [15] (14)

$SiO_2(g)=SiO(g)+O$ [17] (15)

$Hg(g)+Te(l)=HgTe(s)$ [18] (16)

$2Hg(g)+Te_2(g)=2HgTe(s)$ [18] (17)

The heat of these reactions is given in Table 2.

One sees that heats of chemical reactions obey the present theory, not the traditional one. Heats of physical processes obey traditional thermodynamics.

## References

1. I.A. Stepanov: *DEP VINITI*, No 37-B96, (1996). Available from VINITI, Moscow.
2. Stepanov: *DEP VINITI*, No 3387-B98. (1998). Available from VINITI, Moscow.
3. I.A. Stepanov: *7th European Symposium on Thermal Analysis and Calorimetry.* Aug. 30 - Sept. 4. Balatonfuered, Hungary. 1998. Book of Abstracts. P. 402-403.
4. I.A. Stepanov: *The Law of Conservation of Energy for Chemical Reactions.*- http://ArXiv.org/physics/0010052.
5. I. A. Stepanov. The Heats of Reactions. Calorimetry and Van't-Hoff. 1.- http://ArXiv.org/abs/physics/0010054.
6. I. A. Stepanov. The Heats of Dilution. Calorimetry and Van't-Hoff. - http://ArXiv.org/abs/physics/0010075.
7. T. Hill. *An Introduction to Statistical Thermodynamics*, Dover Publications, New York, (1986).
8. A. D. Rusin. Vestnik MGU. Ser. Khim. 1998. V. 39. No 1, 25-29. (www.chem.msu.su:8081/rus/vmgu/welcome.html).
9. I. Barin: *Thermochemical Data of Pure Substances*, VCH, Weinheim, (1989).
10. A. D. Rusin. Vestnik MGU. Ser. Khim. 1998. V. 39. No 3, 147-152.
11. G. Semenov, K. Frantseva, E. Shalkova, Vestnik LGU, Fiz.-Khim., 1970, v. 16, No 3, 82.
12. L. V. Gurvich, I.V. Veitz, et al: *Thermodynamic Properties of Individual Substances*, 4th edn., Hemisphere Pub Co, NY, L., (1989 etc.).
13. E. Kazenas, D. Chizhikov, Yu. Vasjuta, Zh. Fiz. Khim., 1975, v. 49, No 12 [Russ. J. Phys. Chem.].
14. M. V. Gorbachova, et al, Zh. Fiz. Khim., 1998, 72, 3, 416-420 [Russ. J. Phys. Chem.].

15. R. H. Lamoreaux, D. L. Hildenbrand, L. Brewer, J. Phys. Chem. Ref. Data. 1987. V. 16. N 3. 419-445.

16. L. V. Gurvich, V.S. Iorish, et al: *IVTANTHERMO - A Thermodynamic Database and Software System for the Personal Computer. User's Guide*, CRC Press, Inc., Boca Raton, (1993).*

17. S. Shornikov, I. Archakov, M. Shults, Russ. J. Gen. Chem. 2000, v. 70(3), 360.

18. A. Nasar, M. Shamsuddin, J. Less-Common Metals, 1990, v. 161, 87-92.

* I used data from the Russian version of this database on www.openweb.ru/windows/thermo/index_eng.htm

**Table 1**

The heat of some processes measured by the Van't-Hoff equation and by calorimetry

| $Al_2Br_6(g)=2AlBr_3(g)$ [8] | | | |
|---|---|---|---|
| **T, K** | **$\Delta H^{*0}$ (kJ/mol) (8)** | **$\Delta H^{*0}-P\Delta V^0$ (kJ/mol)** | **$\Delta Q$ (kJ/mol) [9]** |
| 500 | 116,98 | 112,83 | 113,32 |
| 600 | 116,98 | 111,9 | 111,76 |
| 700 | 116,98 | 111,16 | 110,17 |
| 800 | 116,98 | 110,33 | 108,57 |
| 900 | 116,98 | 109,50 | 106,95 |
| **$AlBr_3(l)=AlBr_3(g)$ [10]** | | | |
| **T, K** | **$\Delta H^{*0}$ (kJ/mol)** | **$\Delta H^{*0}-P\Delta V^0$ (kJ/mol)** | **$\Delta Q$ (kJ/mol) [9]** |
| 400 | 85,68 | - | 86,19 |
| 500 | 82,72 | - | 81,72 |
| **$2AlBr_3(l)=Al_2Br_6(g)$ [10]** | | | |
| **T, K** | **$\Delta H^{*0}$ (kJ/mol)** | **$\Delta H^{*0}-P\Delta V^0$ (kJ/mol)** | **$\Delta Q$ (kJ/mol) [9]** |
| 400 | 60,06 | 56,74 | 57,5 |
| 500 | 53,42 | 49,27 | 50,12 |

## Table 2

The heat of some chemical reactions measured by the Van't-Hoff equation and by calorimetry

| **Reaction** | **T, K** | **$\Delta H^{*0}$ (kJ/mol)** | **$\Delta H^{*0}$-$P\Delta V^0$ (kJ/mol)** | **$\Delta Q$ (kJ/mol)** |
|---|---|---|---|---|
| $2V_2O_5$(s)=$V_4O_{10}$(g) [11] | 1300 | 113,01 [11] | 102,21 | 105,05 [12] |
| | 1400 | 113,01 [11] | 101,38 | 99,35 [12] |
| $2SnO_2$(s)=2SnO(g)+$O_2$ [13] | 1300 | 1184,1 [13] | 1151,7 | 1151,4 [12] |
| | 1400 | 1184,1 [13] | 1149,2 | 1145,5 [12] |
| 2HgO(g)=2Hg(g)+$O_2$ [14] | 500 | 45,91 [14] | 41,76 | 39,74 [9] |
| | 600 | 45,91 [14] | 40,92 | 39,90 [9] |
| | 700 | 45,91 [14] | 40,09 | 40,05 [9] |
| | 800 | 45,91 [14] | 39,26 | 40,21 [9] |
| $SiO_2$(s)=SiO(g)+O [15] | 1700 | 1045,0 [15] | 1016,8 | 1022,3 [16] |
| | 1800 | 1045,0 [15] | 1015,1 | 1019,9 [16] |
| | 1900 | 1045,0 [15] | 1013,4 | 1015,3 [16] |
| $SiO_2$(g)=SiO(g)+O [17] | 1700 | 458,1 [17] | 443,97 | 448,40[1] |
| | 1800 | 458,1 [17] | 443,14 | 444,96[1] |
| | 1900 | 458,1 [17] | 442,31 | 441,44[1] |
| Hg(g)+Te(l)=HgTe(s) [18] | 800 | -113,91 [18] | -107,26 | -108,30 [9] |
| | 900 | -113,91 [18] | -106,43 | -108,16 [9] |
| 2Hg(g)+$Te_2$(g)= 2HgTe(s) [18] | 800 | -345,49 [18] | -325,55 | -329,51 [9] |
| | 900 | -345,49 [18] | -323,05 | -325,73 [9] |

[1] Data for SiO and O are from [16], H($SiO_2$(g))=H($SiO_2$(s), [16])+ΔH(evaporation, [15])